\documentclass[twocolumn,showpacs,superscriptaddress,preprintnumbers,floatfix]{revtex4}
\usepackage{latexsym}
\usepackage{amsmath}
\usepackage{amsfonts}

\usepackage{graphicx}
   \def\be{\begin{equation}}
   \def\ee{\end{equation}}
   \def\ba{\begin{eqnarray}}
   \def\ea{\end{eqnarray}}

\begin{document}
\addtolength{\belowdisplayskip}{-0.3cm}       
\addtolength{\abovedisplayskip}{-0.3cm}       
\title{Asymptotic Safety in Einstein Gravity and Scalar-Fermion Matter}
\author{G. P. Vacca} 
\affiliation{INFN, sezione di Bologna, via Irnerio 46, I-40126 Bologna, Italy}
\author{O. Zanusso} 
\affiliation{
SISSA, via Bonomea 265, I-34136 Trieste, Italy and INFN , sezione di Trieste, Italy}

\begin{abstract}
{\bf Abstract.}
Within the functional renormalization group approach we study the effective
QFT of Einstein gravity and one self-interacting scalar coupled to $N_f$ Dirac fermions. 
We include in our analysis the matter anomalous dimensions induced by all the
interactions and analyze the highly non linear beta functions determining the
renormalization flow. We find the existence of a non trivial fixed point
structure both for the gravity and the matter sector, besides the usual
gaussian matter one.
This suggest that asymptotic safety could be realized
in the gravitational sector \emph{and} in the standard model.
Non triviality in the Higgs sector might involve gravitational interactions.

\end{abstract}
\maketitle
\section{Introduction}
\vspace{-0.4cm}
The Einstein theory of gravitational interactions appears to be strongly
coupled at Planck scale and is well known to be perturbatively non
renormalizable~\cite{nonren}.
Therefore, as a quantum theory, is generally considered to be an effective
one, nevertheless useful to compute quantum corrections~\cite{donoghue} induced
by graviton exchanges in low energy processes.
At high energies it is generally believed to derive from a fundamental counterpart
such as string theory or to be possibly described in terms of spin foams or
lattice models.
There have been also proposal that gravity may be induced~\cite{sakarov} by other degrees of
freedom, e.g. matter ones, and recently that gravitational field equations may
be derived as thermodinamic properties~\cite{emergence}.
Even in this latter scenario where gravity is considered emergent we expect
the presence, apart from matter quantum fluctuations, 
of quantum fluctuations of the gravitational degrees of freedom~\cite{PV},
which may therefore be described in some domain by an effective QFT.

The Higgs sector of the standard model is also regarded as an effective field
theory since it cannot be
arbitrarily extended to ultraviolet (UV) energies.
This is related
to the triviality problem which manifests itself even in
perturbation theory due to the appearence of a Landau pole when studying the
renormalized self-interaction of the Higgs field.

In the context of QFT there is nevertheless the possibility to have an UV
complete description of the gravitational and standard model theories,
so that they could be considered {\it fundamental}. The paradigm of {\it asymptotic
safety} has been introduced to provide such a situation, which is realized whenever the
renormalization group (RG) analysis of the theory reveals the existence of a fixed
point (FP) with a finite number of UV attractive (relevant) directions in the
theory space~\cite{AS}.

Analytical tools for the investigations of RG flows of QFT models in the
Wilsonian sense~\cite{wilson}, leading to functional renormalization group
equations~\cite{polchinski,wetterich} are available.
In particular increasing evidence that gravity might be asymptotically safe
(and therefore a meaningful QFT) comes from the
analysis of effective average action (EAA) truncations
which include terms associated to the Einstein-Hilbert and cosmological constant~\cite{reuter},
and then scalar matter minimally or non minimally coupled \cite{PP2,PNR},
quartic and higher derivative
terms of the metric~\cite{CP, CPR, BMS}, free matter of all
spin less than two~\cite{DP,PP1,CPR}.

On the other hand, the possibility that asymptotic safety can solve
the triviality as well as the hierarchy problem in the standard model and in
particular in the Higgs sector has been recently explored. Several models for
interacting scalar-fermion matter have been considered, starting from the most
simple ones, both in the symmetric and in the SSB phases~\cite{gies}. The
simplest model does not show the possibility to have non trivial FPs but
for models of increasing complexity there exists a range of parameters suitable
for asymptotic safety to be realized. 

In this letter we report and discuss the results of a study of Einstein
gravity interacting with the simplest possible system with one
scalar field and $N_f$ Dirac fermions with a $U(N_f)$ symmetric Yukawa
interaction, which by itself has been found to be a non asymptotic safe
theory. The investigations on such a system started in \cite{ZZVP}.
For a pertubative approach at one loop with dimensional regularization see also~\cite{rodigast}.

The goal is twofold. On one hand we would like to see if the inclusion of
such an interacting matter can spoil the formation of the non trivial FP
structure in the gravity sector. In doing this we also consider for the first time the
anomalous dimensions of the matter fields, which have been previously
neglected in all asymptotic safety studies which include gravity.
On the other hand we want to analyze if in this simplified model the
gravitational interaction leads to the possibility to have an asymptotically
safe matter sector, eventually removing the triviality problem in the
scalar sector.

This might serve as a prelude for a more general analysis involving the Higgs,
fermion and gauge fields of the standard model. There is the
possibility that the UV completion of such a theory together with gravity might
be realized in a satisfactory way in the QFT sense. This would constitute an
alternative to considering a scenario were these theories emerge at
low energy from superstring theories or other viable GUT.
\section{Gravity and Yukawa system}
\vspace{-0.4cm}

We adopt Wilson's approach to renormalization
and introduce an euclidean EAA with a scale dependence.
We truncate it to the form
\ba
 \Gamma_k\left[\Upsilon\right]
 & = & \int \mathrm{d}^4 x \sqrt{g} (L_b+L_f+L_g)
\ea
where the single terms are
\ba
 L_b &=& \tfrac{1}{2}Z_\phi\nabla^\mu\phi\nabla_\mu\phi+V(\phi)
 \label{scalar action}\\
 L_f &=& \tfrac{i}{2}Z_\psi(\bar{\psi}\gamma^\mu D_\mu\psi-D_\mu\bar{\psi}\gamma^\mu\psi)
         +i\,H(\phi)\,\bar{\psi}\,\psi)
 \label{spinor action}\\
 L_g &=& - Z \sqrt{g}R \left[g_{\mu\nu}\right]
 \label{gravitational action}
\ea
The multiplet $\Upsilon^A=(\phi,\psi,\bar{\psi},g_{\mu\nu})$ contains all the dynamical fields.
$k$ is a sliding renormalization scale that parameterizes the RG flow of the
EAA. The Newton constant is contained in $Z=1/16\pi G$, while the
cosmological constant
is the zero point energy of the potential $V\left(\phi\right)$.
$H\left(\phi\right)$ is a generalized Yukawa interaction.
The spinor covariant derivative is
$D_\mu\psi=\partial_\mu\psi+1/2\omega_{\mu c d}J^{c d}\psi$,
where $\omega_{\mu c d}$ is the spin connection and 
$J^{cd}=1/4[\gamma^c,\gamma^d]$ are the $O(4)$ generators.

We use the background field method in conjunction to exact renormalization group.
The action is expanded around a background $\Upsilon$ using fluctuations
$\Sigma^{\hat{A}}=(\varphi,\chi,\bar{\chi},h_{\mu\nu},\bar{c}_\mu,c_\mu)$ and define
\ba
 \varGamma_k\left[\Sigma,\Upsilon\right]
 &=& \Gamma_k\left[\Upsilon+\Sigma\right]+ L_{g.f.}+L_{gh}
\ea
We add a gauge fixing term for the diffeomorphisms and a corresponding ghost action
\ba
&&
L_{g.f.}\!=\!\frac{Z}{2\alpha} g^{\mu\nu}F_\mu F_\nu\ ;\
F_\mu\!=\!\left(\delta_\mu^\beta \nabla^\alpha
-\frac{1\!+\!\beta}{4}g^{\alpha\beta}\nabla_\mu\right) h_{\alpha\beta}
\nonumber\\
&&
L_{gh}=\bar{c}_\mu \left(-g^{\mu\nu}\nabla^2
+\frac{\beta-1}{2}\nabla^\mu \nabla^\nu-R_{\mu\nu}
\right)c_\nu\ .
\ea
Notice that we included the ghosts in the fluctuations multiplet.
The $O(4)$ gauge is fixed such that the vierbein is symmetric and
all their fluctuations can be written in terms
of $h_{\mu\nu}$. There are no $O(4)$ ghosts.

The coarse-graining procedure is achieved using an IR cutoff term
\ba
\Delta S_k &=& \int \mathrm{d}^4 x \sqrt{g} \Sigma^{\rm T} {\cal R}_k\left[\Upsilon\right] \Sigma
\ea
We define the modified propagator
${\cal G}_k \equiv (\varGamma_k^{(2,0)}\left[\Upsilon,0\right]+{\cal R}_k\left[\Upsilon\right])^{-1}$.
The apex numbers $(2,0)$ indicate that two functional derivatives w.r.t. the first argument were taken.
The flow of the EAA is governed by the exact RG equation (ERGE)~\cite{wetterich}
\ba
 k\partial_k \Gamma_k
 &=&
 \frac{1}{2} {\rm STr} {\cal G}_k \dot{\cal R}_k 
= \frac{1}{2} \,\,\,\,\,\scalebox{0.15}{\includegraphics[viewport=50 45 120 120]{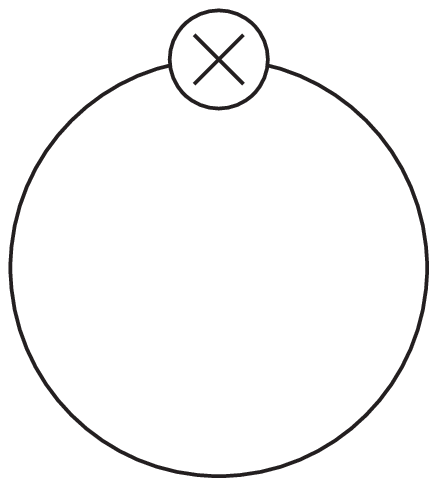}}
\label{ERGE}
\ea
where the dot means $t={\rm Log}\, k/k_0$ derivative and $k_0$ is a reference
scale; $\partial_t=k\partial_k$.
We require the cutoff kernel to be a function of the background metric
$g_{\mu\nu}$ only, i.e. ${\cal R}_k={\cal R}_k\left[g\right]$.
In particular, the cutoff is required to be a function of the operators
$\Box=-\nabla^2$ and $i\gamma^\mu D_\mu$, so that any covariant derivative
appearing in $\varGamma_k^{(2,0)}\left[\Upsilon,0\right]$ is coarse-grained by
${\cal R}_k={\cal R}_k\left[g\right] $. 
The profile function of the cutoff is always the optimized one~\cite{litim},
$R_{k,{\rm opt.}}(z)=(k^2-z)\theta(k^2-z)$, for $z=\Box$, and allows to perform exacty the trace.
The choice we use is often referred as ``type-I'' cutoff in the existing
literature \cite{CPR} and coincides with that of \cite{DP}.
Other cutoff choices are possible and we will comment more on them in the discussions.

The flow of the functions $V(\phi)$ and $H(\phi)$ is calculated using
(\ref{ERGE}) evaluated in a background for which
$\phi={\rm const.}$, $\psi={\rm const.}$ and
$g_{\mu\nu}=\delta_{\mu\nu}$. All this, apart for a slightly different choice of the
cutoff scheme, goes along the lines of our previous work~\cite{ZZVP}.
In order to calculate the flow of $Z_\phi$ and $Z_\psi$, we
take two functional derivatives of (\ref{ERGE}) with respect to
$\Upsilon^A$ and $\Upsilon^B$ such that $\Upsilon^A,\Upsilon^B\neq
g_{\mu\nu}$.
We obtain the flow of the $2$-point function
\ba
k\partial_k \frac{\delta^2 \Gamma_k}{\delta\Upsilon^A\delta\Upsilon^B}
&=&
   {\rm STr}
   {\cal G}_k
   \frac{\delta \varGamma_k^{(2,0)}}{\delta\Upsilon^{(A}}
   {\cal G}_k
   \frac{\delta \varGamma_k^{(2,0)}}{\delta\Upsilon^{B)}}
   {\cal G}_k
   \dot{\cal R}_k
\nonumber\\
& &
   -\frac{1}{2}{\rm STr}
   {\cal G}_k
   \frac{\delta^2 \varGamma_k^{(2,0)}}{\delta\Upsilon^{A}\delta\Upsilon^{B}}
   {\cal G}_k
   \dot{\cal R}_k
\nonumber\\
&=&
 \,\,\,\,\scalebox{0.15}{\includegraphics[viewport=40 50 120 120]{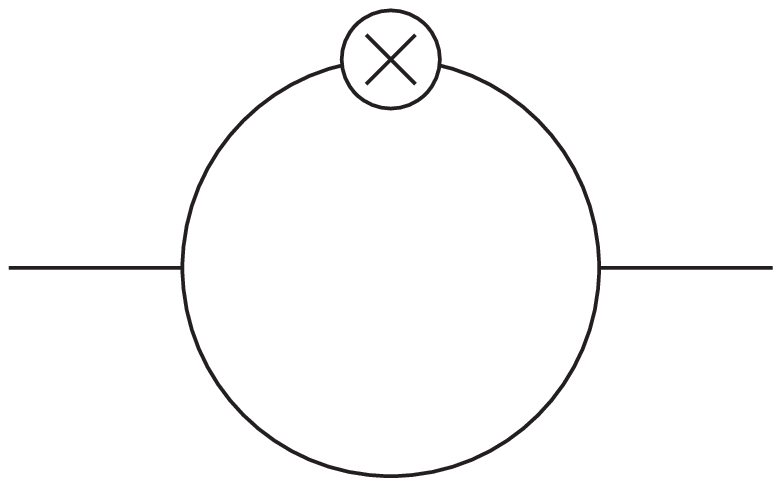}}
 \,\,\,\,\,\,\,\,\,\,\,\,-\frac{1}{2}
 \,\,\,\,\,\scalebox{0.15}{\includegraphics[viewport=50 45 120 120]{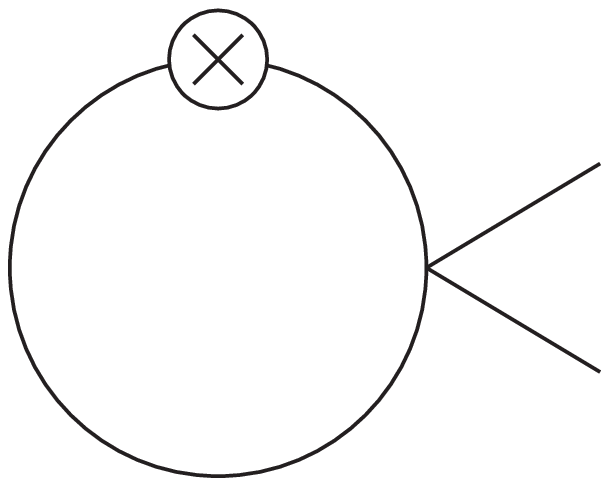}}
\label{ERGE2}
\ea
Matrix indices have been omitted for notational simplicity.
The derivatives did not act on ${\cal R}_k$ because it is a function of the metric only.
The anomalous dimensions are defined as $\eta_\phi=-\dot{Z}_\phi/Z_\phi$ and 
$\eta_\psi=-\dot{Z}_\psi/Z_\psi$. We use (\ref{ERGE2}) in the background
$\phi={\rm const.}$, $\psi=0$, $g_{\mu\nu}=\delta_{\mu\nu}$ and
then move to momentum space. Additionally, we require $V'(\phi)=0$ to simplify to some extent.
Choosing in (\ref{ERGE2}) $\Upsilon^A=\Upsilon^B=\phi$, or $\Upsilon^A=\psi$ and
$\Upsilon^B=\bar{\psi}$,
we calculate $\eta_\phi$ and $\eta_\psi$, isolating the coefficients of the
operators $Z_\phi \Box$ and $i Z_\psi \gamma^\mu D_\mu$, respectively.
$\eta_\phi$ and $\eta_\psi$ calculated in this way are \emph{on-shell} and have a $\phi$ dependence
that has to be fixed by physical requirements.

The running $\dot{Z}$ is calculated using heat kernel methods.
We set the background to be a $4$-sphere with corresponding metric
and choose $\phi={\rm const.}$ and $\psi=0$.
(\ref{ERGE}) is then written as the trace of a function of $\Box$.
The Seeley-DeWitt expansion gives the term linear in
$R\left[g_{\mu\nu}\right]$ of (\ref{ERGE}), 
from which we read the beta function $\dot{Z}$.
\vspace{-0.5cm}
\section{RG flow}
\vspace{-0.3cm}
Using (\ref{ERGE2}) and the heat kernel technique, we get a system for the
quantities $\{\eta_\phi,\eta_\psi,\dot{\bar{Z}}\}$ which has to be solved.
$\bar{Z}=k^{-2}Z $ is the dimensionless Planck mass.
$\eta_\phi$, $\eta_\psi$ and $\dot{\bar{Z}}$ depend on the constant 
field configuration $\phi$ through the functions $H$ and $V$.
We set it to be $\left<\phi\right>$ in order to agree with the initial truncation
(\ref{scalar action},\ref{spinor action},\ref{gravitational action}) that 
has constant $Z_{\phi,\psi}$. From (\ref{ERGE}) we calculate $\dot{V}$ and $\dot{H}$.
It is convenient to use the dimensionless renormalized field
$\phi_R=Z_\phi^{1/2} k^{-1} \phi$ and functions
$ v[\phi_R]=k^{-4} V[\phi]$, $h[\phi_R]=k^{-1} Z_\psi^{-1} H[\phi] $. 
We obtain a functional system
of the form
\ba
\dot{v}&=&F_1(v,v',v'',h,\bar{Z},\dot{\bar{Z}},\eta_\phi,\eta_\psi)\\
\dot{h}&=&F_2(h,h',h'',v',v'',\bar{Z},\dot{\bar{Z}},\eta_\phi,\eta_\psi)
\ea
If we substitute the previous relations for $\{\eta_\phi,\eta_\psi,\dot{\bar{Z}}\}$,
anomalous dimensions are eliminated and we remain
with the system $\{\dot{v},\dot{h},\dot{\bar{Z}}\}$. The details of
these partial differential equations, and of
$\dot{\bar{Z}}$, $\eta_\phi$ and $\eta_\psi$ will be given elsewhere~\cite{VZ}.

Now we truncate further $v$ and $h$ to get a
finite system of ordinary differential equations. We
adopt a power expansion of $v$,
that is valid for values of $\phi_R$
within a scale dependent radius of convergence, to some finite order.
A $\mathbb{Z}_2$-symmetric phase is analyzed through the expansion
$v = \lambda_0+\lambda_2\phi_R^2+\lambda_4\phi_R^4$ where
$\left<\phi_R\right>=0$ and $\lambda_2>0$.
Alternatively, if $\lambda_2<0$, we are in a symmetry breaking regime and expand
$v = \theta_0+\theta_4(\phi_R^2-\kappa)^2$. The crucial difference of
the latter regime is that the  running of the VEV
$\left<\phi_R\right>=\sqrt{\kappa}$ is determined by the requirement $v'(\sqrt{\kappa})=0$.
In general the VEV beta function acquires gravitational correction only through
the anomalous dimensions of matter fields:
\vspace{0.1cm}
\ba
 \dot{\kappa}\!=\!\!
 -\!\left(2\!+\!\eta_\phi\right)\! \kappa
 \!+\!\frac{v^{\prime\prime\prime}\!\left(1\!-\!\frac{\eta_\phi}{6}\right)\!\!\sqrt{\kappa}}
       {16\pi^2 v^{\prime\prime}\!\left(1\!+\!v^{\prime\prime}\right)^2}
 \!-\!\frac{N_{\!f}h h^{\prime} \!\!\left(1\!-\!\frac{\eta_\psi}{5}\right)\!\!\sqrt{\kappa} }
       {2\pi^2 v^{\prime\prime}\!\left(1\!+\!h^2\right)^2 }\Bigg|_{\bar{\phi}_R=\sqrt{\kappa}}
\nonumber
\ea
We further consider a simple Yukawa interaction according to
$h(\bar{\phi}_R)=y\, \bar{\phi}_R$. In both phases, with truncations
up to fourth order in the scalar potential, the RG flow is encoded in a system
of 5 ODEs. We mainly considered them in the DeDonder gauge
($\alpha=0$, $\beta=1$) but investigated also other choices.

Let us present the FP structure for the {\it symmetric phase}, which depends
on the parameter $N_f$.
In particular we find the expected fully Gaussian (GFP) as well as the Gaussian
matter fixed points (GMFP), that exists for any value of $N_f$. The latter has
$-1< \eta_\psi<0$ and, in the DeDonder gauge,  $\eta_\phi=0$.
The ultraviolet critical surface has dimension $4$ (but $3$ for the fully
Gaussian case for $N_f<3$).

Moreover we find the existence of a fully non trivial FP for $N_f>3$,
with small anomalous dimensions. The dimension of the UV critical
surface is always $3$, independently of $N_f$.
We present in table $1$ few examples of FP and the asymptotic behavior
$N_f\to\infty$. We also show the values of critical exponents $c_i$.
\vspace{-0.4cm}
\begin{widetext}
\begin{center}
\begin{tabular}{|l|l|l|l|l|l|l|l|l|l|l|l|l|}
 \hline
 $N_f$ & $\lambda _0$ & $\lambda _2$ & $\lambda _4$ & $y$ & $\bar{Z}$ & $\eta _{\psi }$ & $\eta _{\phi }$ &$c_1$&$c_2$&$c_3$&$c_4$&$c_5$\\
 \hline
 4 & -0.0129 & 0.243 & 0.832 & 2.70 & 0.00562 & -0.215 & 0.400 &-1.02 & -0.566 & 1.30 & 1.38 & 3.88\\ 
 8 & -0.0260 & 0.0872 & 0.437 & 1.25 & 0.00700 & -0.102 & 0.225 & -0.613 & -0.240 & 1.66 & 1.68 & 3.96 \\
 20 & -0.0642 & 0.0289 & 0.0543 & 0.467 & 0.0130 & -0.0407 & 0.0958 & -0.235 & -0.0937 & 1.86 & 1.87 & 3.99\\
 $N_f \to \infty$ & $\lambda_{0,\infty}-\frac{N_f}{32\pi^2}$&
 $\frac{\lambda_{2,\infty}}{N_f}$ 
& $\frac{\lambda_{4,\infty}}{N_f^2}$&  $\frac{y_{\infty}}{N_f}$&
 $Z_{\infty}+\frac{N_f}{192\pi^2}$ 
& $\frac{\eta_{\psi,\infty}}{N_f}$ & $\frac{\eta_{\phi,\infty}}{N_f}$&
 $\frac{c_{1,\infty}}{N_f}$ & $\frac{c_{2,\infty}}{N_f}$ &
 2+$\frac{c_{3,\infty}}{N_f}$ & 2+$\frac{c_{4,\infty}}{N_f}$ & 4+$\frac{c_{5,\infty}}{N_f}$\\
 \hline
\end{tabular}
\end{center}
Tab. $1$. \,\, Position of the non trivial fixed point, the critical exponents
and the matter field anomalous dimensions for varying $N_f$ and their
asymptotic values. The asymptotic constants have been determined analytically~\cite{VZ}.
\end{widetext}
\vspace{-1cm}
The non trivial fixed point approaches the GMFP in the limit $N_f \to \infty$.
We also considered scalar potential truncations with higher order polinomials.
The shape of the potential at the FP changes very little and
the FP exists with negligible variations for $\lambda_0$, $y$ and
$\bar{Z}$ for all values of $N_f$. Moreover the matter field
anomalous dimensions appear to be small and so we expect
the truncation to be reliable.

There is little sensitivity with respect to the choice of the cufoff
function, as was already discussed in~\cite{ZZVP}. 
Adopting a pure cutoff scheme, where even the field strength of the fields is not
included in ${\cal R}_k$, is leading to very similar results.
Moreover we have observed a very good stability of the results on changing the
gauge parameters $\alpha\ge 0$ and $\beta$.
The dependence on the latter is due to the offshellness of the
flat space background choice and to the truncation employed in \eqref{ERGE}.

We consider now the {\it SSB phase} with a non zero scalar VEV. The numerical
search for the FPs has revealed five $N_f$-dependent families
of solutions. There is a (generalized) GMFP solution ($\theta_4=y=0$ and $\kappa=3/32\pi^2$)
with $\eta_\phi=0$ and tipically $|\eta_\psi|\ll 1$ and UV critical dimension
$4$. Another solution exists for $N_f=1,2$ and has $y=0$, small anomalous
dimensions and is completely UV repulsive.
The other three solutions, completely non trivial,
are characterized by at least one large anomalous dimension and
therefore are not very appealing.
\section{Discussions}
\vspace{-0.3cm}
We have computed, using functional renormalization
group techniques, quantum corrections in a special but non trivial interacting
gravity matter system, taking into account the anomalous dimensions of the
matter fields (one scalar and $N_f$ Dirac fermions).
The results we
obtained corroborate the scenario of an asymptotically safe gravity.
A non trivial FP in the gravitational sector is realized as
GMFP, independently of the cutoff schemes used in both symmetric and broken phases.

Moreover we find that in the symmetric phase another FP exists,
completely non trivial, i.e. with non constant scalar potential $v$
and non zero Yukawa coupling associated to a three dimensional critical
surface. Such a point is stable under different polynomial expansions of the
local scalar potential and change of the gauge fixing parameters.
We
found that, when neglecting the anomalous dimensions,
this FP exists for $5<N_f\le12$, with couplings of the same order of
magnitude, and that another non trivial one appears, with much larger values
and a different $N_f$ asymptotic dependence.
The lesson we learn is that, in general, anomalous dimensions should be taken
into account as this might be crucial to determine the correct FP structure.
In the broken phase several non trivial FP solutions appears,
but due to their large anomalous dimensions we do not consider them
physically trustable.

In the previous sections we studied in detail the RG flow adopting a
type-I cutoff scheme. In such a scheme, for $n_s$ scalar, $n_D$ Dirac fermions
and $n_M$ massless vector fields, the gravitational $\beta$ function
acquires a leading term proportional to $(-n_s+n_D+\frac{7}{4}n_M)$
\cite{CPR}.
Another often used scheme (type-II)
is such that the same contribution is proportional to
$(-\frac{1}{2}n_s-n_D+2n_M)$.
The contribution to the vacuum energy is the same, proportional
to $(n_s-4n_D+2n_M)$ in both schemes \cite{CPR}.
Studying the RG flow of our simple model using the type-II cutoff
one finds that, for any $N_f$, the non trivial FP disappears.
However it is clear that this feature may change when introducing a more realistic matter sector.
As a preliminary test we
considered our
simplified toy model with additional free scalar, Dirac fermion and vector fields.
Having in mind to keep a Yukawa coupling for the top quark only,
which belongs to a fundamental representation of ${\rm SU}(3)_c$,
we set  $N_f=3$ and the number of the other free fields
to match the content of the standard model. Repeating the RG flow analysis, we
found for both cutoff type schemes (I and II) a non trivial FP.
This FP is non trivial, both in the gravitational and in the
interacting scalar-fermion sector and has a Yukawa coupling with value about
unity. The non trivial FP exists also when changing
the number of the extra free fields around the values corresponding to the
standard model content, for a wide range of parameters.
We observe that tipically it has a negative value $\lambda_0$, which
nevertheless may change sign in the RG flow towards the IR. 
An analysis of the influence of free fields on the gravitational FP
with type-I cutoff was previously presented in~\cite{PP1}.
Details of our analysis will be given in~\cite{VZ}.

In a scenario with an asymptotically safe gravity,
there has been a proposal to predict the Higgs boson mass~\cite{SW}.
Negative gravitational corrections to the top Yukawa beta function were
required. However, if we expand perturbatively our results and retain
the leading order, we find a negative gravitational contribution to the
fermion anomalous dimensions, but a positive
gravity-induced term to the Yukawa beta function, $\beta_y^{grav}=a_y y \,G
k^2$ with $a_y\sim 0.2$. 

It is intriguing that in the symmetric phase we get a fully non trivial FP
solution with a Yukawa coupling value around unity,
which is close to the top Yukawa one at low energies.
One may imagine a scenario with a modest flow of the Yukawa coupling
from UV (ultra Planckian) to IR (Fermi) scales.
Nevertheless, one should include a mechanism to obtain a low energy SSB phase.
On the other hand a full investigation of a situation closer to the standard
model might reveal the possibility that a trustable non trivial fixed point
is realized even in an Higgs SSB phase.
We plan to study and discuss more these issues elsewhere.

{\bf Acknowledgements}: The authors wish to thank R. Percacci for critical
reading of the manuscript and A. Codello for useful discussions.
\vspace*{-0.4cm}

\end{document}